# Pure Bulk Orbital and Spin Photocurrent in Two-Dimensional Ferroelectric Materials

Xingchi Mu, Jian Zhou[*]

*Center for Advancing Materials Performance from the Nanoscale, State Key Laboratory for Mechanical Behavior of Materials, Xi'an Jiaotong University, Xi'an 710049, China*

**Abstract**

We elucidate light-induced orbital and spin current through nonlinear response theory, which generalizes the well-known bulk photovoltaic effect in centrosymmetric broken materials from charge to the spin and orbital degrees of freedom. We use two-dimensional nonmagnetic ferroelectric materials (such as GeS and its analogues) to illustrate this bulk orbital/spin photovoltaic effect, through first-principles calculations. These materials possess a vertical mirror symmetry and time-reversal symmetry but lack of inversion symmetry. We reveal that in addition to the conventional photocurrent that propagates parallel to the mirror plane (under linearly polarized light), the symmetric forbidden current perpendicular to the mirror actually contains electron flows, which carry angular momentum information and move oppositely. One could observe a pure orbital moment current with zero electric charge current. This hidden photo-induced orbital current leads to a pure spin current via spin-orbit coupling interactions. Therefore, a four-terminal device can be designed to detect and measure photo-induced charge, orbital, and spin currents simultaneously. All these currents couple with electric polarization $P$, hence their amplitude and direction can be manipulated through ferroelectric phase transition. Our work provides a route to generalizing nanoscale devices from their photo-induced electronics to orbitronics and spintronics.

**Introduction.** Bulk photovoltaic (BPV) effect,[1] which converts incident alternating optical field into direct electric current in centrosymmetric broken materials, has attracted tremendous attention during the past few decades for its easy manipulation and low energy cost. Comparing with conventional light-to-current conversion in a p-n junction between two semiconductors, BPV effect produces electric current everywhere light shines onto the material, which could significantly enhance the conversion efficiency and density. From physics point of view, BPV effect is a second order nonlinear optical effect, which includes two photons (with frequency $\omega$ and $-\omega$; absorption and emission) and an electron (with moving velocity $v$).[2]

The BPV effect uses electron charge degree of freedom (DOF) to generate a biased electric potential in semiconductors,[3-6] which are serving as promising electronic devices. In order to further increase the information read/write kinetics and storage density, one may resort to new DOFs of electron, such as its spin angular momentum. The study of the intrinsic spin and its induced magnetic moment is thus referred to as "spintronics",[7-9] which has been shown to hold unprecedented potential in the future miniaturized devices, especially in the field of quantum computing and neuromorphic computing. Roughly speaking, when the velocities of electrons in the spin up and spin down channels are different ($v_\uparrow - v_\downarrow \neq 0$), there is a collection motion of electron spin and leads to a nonzero spin current.

In addition to spin, another DOF that could produce angular momentum and magnetic moment is the electron orbital which describes the electron travelling around one or a few nuclei. It is an overlooked DOF because in most conventional materials, the orbital moment is significantly or completely quenched under strong and symmetric crystal field. However, when the symmetry and strength of crystal field are reduced, especially in low-dimensional materials, orbital DOF may play an important role in their magnetic properties, topological behaviors, and valleytronic features.[10-13] Similar as spintronics, this novel field is thus termed as "orbitronics",[14,15] which is predicted to further enhance information read/write speed significantly. If the electron velocities carrying different orbital magnetic moments are different (e.g., $v_{l_z} - v_{-l_z} \neq 0$), then one could also expect an *orbital current*, analogous to spin current. Note that such an orbital current has been predicted in the linear response Hall effect picture,[16-18] in addition to spin Hall effect[19,20] and valley Hall effect.[11,21-23]

In the current work, we predict that in addition to nonlinear BPV effect, there exists a *hidden* orbital current which carries colossal orbital moment when light shines onto two-dimensional (2D) nonmagnetic ferroelectric materials. We refer to this effect as bulk orbital photovoltaic (BOPV) effect, which is described by a second order nonlinear optical process. We use 2D ferroelectric group-IV monochalcogenide monolayers (GeS, SnS, GeSe, SnSe, GeTe, and SnTe)[24-28] and group-V single elemental monolayer Bi(110)[29,30] to illustrate our theory. This family of materials have been experimentally fabricated and proved to possess flexible and robust in-plane ferroelectrics.[31,32] They all belong to $Pmn2\bar{1}$ space group, which contains a vertical mirror symmetry ($\widehat{\mathcal{M}}_x$), and the electric polarization is along $y$. When linearly polarized light (LPL) (with their polarization direction along $x$ or $y$) is irradiated onto them, conventional nonlinear BPV current is parallel to the polarization direction $y$, while the net BPV along $x$ is zero, according to symmetry considerations. However, we apply first-principles calculations and reveal that there are unexpected hidden electron movements along the $+x$ and $-x$, which carries different orbital moments. Hence, one could expect a finite and pure BOPV current to be measured in the $x$-direction. Here pure orbital current means no electric charge current is mixed. We also show that the spin-orbit coupling (SOC) interaction could convert such BOPV current into spin DOF, namely, bulk spin photovoltaic (BSPV) current (also along $x$).[33-36] When circularly polarized light (CPL) is used, the conventional charge BPV current will be along $x$, while the BOPV and BSPV currents flow along $y$.

**Results and Discussion.** When light (propagating along out-of-plane direction $z$) irradiates onto the 2D materials, one could adopt closed circuit boundary condition[37] and use electric field ($\boldsymbol{E}$) as the natural variable. The BPV, BOPV, and BSPV effect can be evaluated by (Einstein summation convention adopted)

$$\mathcal{J}^c(\omega = 0) = \sigma^c_{ab}(0; \omega, -\omega) E_a(\omega) E_b(-\omega),$$

$$\mathcal{J}^{c;L_i}(\omega = 0) = \sigma_{ab}^{c;L_i}(0;\omega,-\omega)E_a(\omega)E_b(-\omega),$$

$$\mathcal{J}^{c;S_i}(\omega = 0) = \sigma_{ab}^{c;S_i}(0;\omega,-\omega)E_a(\omega)E_b(-\omega). \quad (1)$$

Here $\mathcal{J}^c$, $\mathcal{J}^{c;L_i}$, $\mathcal{J}^{c;S_i}$ are electric charge current, orbital current, and spin current density that propagate along the $c$-direction, respectively, and $E_a$ ($E_b$) is electric field component ($a,b,c = x,y$). $L_i$ and $S_i$ are orbital and spin angular momentum components, respectively. Here, we focus on their $z$-component ($i = z$) which is most frequently measured and observed experimentally. Actually we have demonstrated that under glide plane symmetry $\{\widehat{\mathcal{M}}_z | \left(\frac{1}{2}\frac{1}{2}0\right)\}$, the in-plane spin angular momentum component is suppressed in most $k$-points of the Brillouin zone (BZ).[30] Now we focus on the $\widehat{\mathcal{M}}_x$ mirror plane effects. One can apply a simple symmetry analysis to examine the non-dissipation response features of these coefficients. Under LPL ($a = b = x,y$), which does not break $\widehat{\mathcal{M}}_x$ symmetry, the charge current would also obey $\widehat{\mathcal{M}}_x$. Hence, $\mathcal{J}^x = \widehat{\mathcal{M}}_x \mathcal{J}^x = -\mathcal{J}^x = 0$, so that both $\sigma_{xx}^x$ and $\sigma_{yy}^x$ are zero (symmetry forbidden). On the other hand, the vertical direction current $\mathcal{J}^y$ can be nonzero, hence the $\sigma_{xx}^y$ and $\sigma_{yy}^y$ are finite, indicating that BPV electric charge current only flows along $y$. This is consistent with previous works.[38,39] However, since both $L_z$ and $S_z$ transform as pseudovectors – they flip their sign under $\widehat{\mathcal{M}}_x$. One thus expects that the $\sigma_{xx}^{x;L_z}$, $\sigma_{xx}^{x;S_z}$, $\sigma_{yy}^{x;L_z}$ and $\sigma_{yy}^{x;S_z}$ would become nonzero, while the $\sigma_{xx}^{y;L_z}$, $\sigma_{xx}^{y;S_z}$, $\sigma_{yy}^{y;L_z}$ and $\sigma_{yy}^{y;S_z}$ all vanish. These suggest that the zero valued BPV current $\mathcal{J}^x$ actually do not indicate the electron motions are completely frozen in the $x$-direction. There exists hidden electron flows, which carries orbital and spin angular momentum instead of charge DOF. We illustrate this hidden spin/orbital photocurrent in Figure 1(a). Since the charge photocurrent and spin/orbital angular momentum (or magnetic moment) photocurrent are perpendicular to each other, one could apply a four-terminal device to measure and observe them simultaneously.

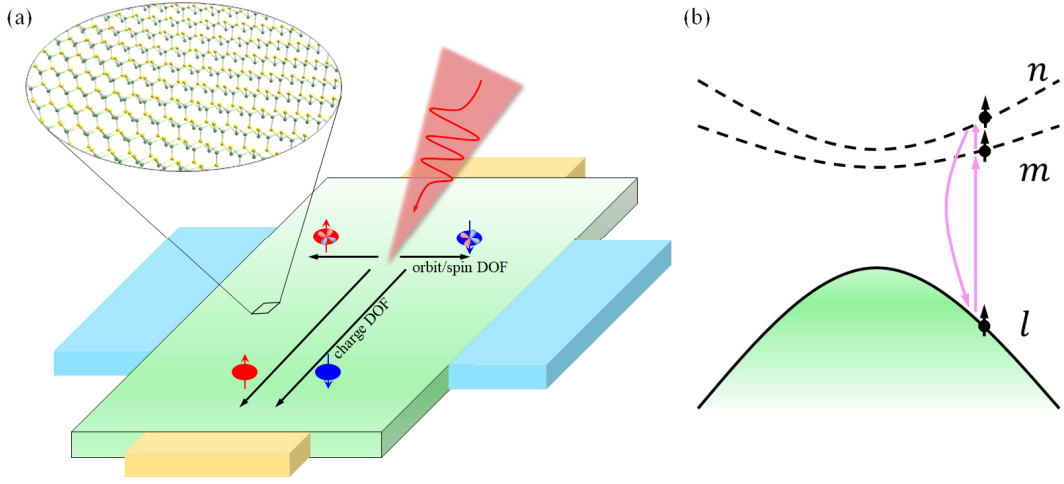

**Figure 1.** (a) Schematic plot of bulk angular momentum photo-conductivity (BOPV and BSPV) effect. Unlike BPV that propagates along the mirror plane of monolayer GeS, the BOPV and BSPV currents carrying orbital and spin DOF information are travelling normal to the mirror plane ($M_x$) under light. A four-terminal device can be used to detect all these photocurrents. The yellow electrode measures charge DOF, while the blue electrodes are magnetic and could measure angular momentum DOF. (b) Interband transition (pink arrows) picture, where black arrows indicate angular momentum or magnetic moment from orbital and spin DOF. Solid and dashed curves represent occupied (valence) and unoccupied (conduction) bands of the semiconductor, respectively.

We calculate the nonlinear photo-conductivity coefficients explicitly. According to the second order Kubo response theory and within the independent particle approximation framework,[5,40] one can compute the complex nonlinear photo-conductivity from its band structure via

$$\chi_{ab}^{c;\mathcal{O}_z}(0;\omega,-\omega) = \frac{e^2}{\hbar^2\omega^2}\int\frac{d^3\mathbf{k}}{(2\pi)^3}\sum_{lmn}^{\Omega=\pm\omega} f_{ln}\frac{v_{nl}^a v_{lm}^b j_{mn}^{c;\mathcal{O}_z}}{(\omega_{nm}-i/\tau)(\omega_{nl}+\Omega-i/\tau)}. \qquad (2)$$

This is based on a three band model that includes band-$m$, $n$, and $l$. The phenomenological carrier lifetime $\tau$ is taken to be 0.2 ps, and $i=\sqrt{-1}$. The direction scripts $a,b,c=x,y$. $f_{ln}=f_l-f_n$ and $\omega_{nl}=\omega_n-\omega_l$ are the occupation and frequency difference between band-$l$ and band-$n$, respectively. Velocity parameter is defined as $v_{lm}^a=\langle l|\hat{v}^a|m\rangle$ and the orbital (and spin) current operator is adopted to be $j_{mn}^{c;\mathcal{O}_z}=\langle m|\{v_c,\mathcal{O}_z\}|n\rangle=\frac{1}{2}\langle m|v_c\mathcal{O}_z+\mathcal{O}_z v_c|n\rangle$ with $\mathcal{O}_z=L_z$ (or $S_z$). For the charge current, it can be replaced by $j_{mn}^c=\langle m|ev_c|n\rangle$. The explicit $\mathbf{k}$-dependence on these quantities are omitted. If during interband transition the $j_{mn}^{c;L_z}$ or $j_{mn}^{c;S_z}$ rises, it could lead to finite BOPV or BSPV current. We schematically plot this physical process in Figure 1b. Note that this is different from the intra-band optical responses in doped semiconductors.[41]

Under LPL, the nonlinear photo-conductivity in Eq. (1) can be evaluated as $\sigma_{aa}^{c;O_i} = \Re \chi_{aa}^{c;O_i}$. For CPL irradiation (along $z$), since the field takes the form $i[\boldsymbol{E}(\omega) \times \boldsymbol{E}^*(\omega)]_z$, the response can be evaluated by $\sigma_{ab}^{c;O_i} = \frac{1}{2}\Im(\chi_{ab}^{c;O_i} - \chi_{ba}^{c;O_i})$. We will focus on LPL photo-conductivity in the main text, and plot and discuss the CPL responses in Supporting Information. We can denote the numerator of Eq. (2 in the format $N_{abc;z} = v^a v^b v^c O_z$, which can be used to determine the (dissipationless) symmetry allowed and forbidden responses. The time-reversal symmetry gives $\hat{\mathcal{T}} v^a(\boldsymbol{k}) = -v^{a,*}(-\boldsymbol{k})$ and $\hat{\mathcal{T}} O_z(\boldsymbol{k}) = -O_z^*(-\boldsymbol{k})$, hence, one could have $\hat{\mathcal{T}} N_{abc;z}(\boldsymbol{k}) = N_{abc;z}^*(-\boldsymbol{k})$ for a Kramers pair (here $\cdot^*$ indicates complex conjugate of quantity $\cdot$). Integration over the first Brillouin zone (BZ) yields a real number. As for mirror symmetry, since $\hat{\mathcal{M}}_x v^x(\boldsymbol{k}) = -v^x(\tilde{\boldsymbol{k}})$, $\hat{\mathcal{M}}_x v^y(\boldsymbol{k}) = v^y(\tilde{\boldsymbol{k}})$ ($\tilde{\boldsymbol{k}}$ is mirror symmetric image of $\boldsymbol{k}$, $\tilde{k}_x = -k_x, \tilde{k}_y = k_y$), and $\hat{\mathcal{M}}_x O_z(\boldsymbol{k}) = -O_z(\tilde{\boldsymbol{k}})$, we have $\hat{\mathcal{M}}_x N_{aax;z}(\boldsymbol{k}) = N_{aax;z}(\tilde{\boldsymbol{k}})$ and $\hat{\mathcal{M}}_x N_{aay;z}(\boldsymbol{k}) = -N_{aay;z}(\tilde{\boldsymbol{k}})$. The latter is odd under mirror operation. We thus prove that BOPV and BSPV currents only occur along the $x$-direction under $x$ or $y$-LPL, consistent with previous analysis. In the long relaxation time approximation, one could demonstrate that LPL irradiation yields the BPV shift current and the CPL illumination gives an injection current for time-reversal symmetric systems. However, for the BSPV and BOPV, we find that the LPL induced photocurrent is injection-like, which is proportional to the relaxation time $\tau$. The CPL, on the other hand, induces shift-like current (see Supporting Information for detailed discussions).[42]

Now we apply Eq. (2) to compute nonlinear photo-conductivity in 2D nonmagnetic ferroelectric materials. Taking monolayer GeS as an example (Figure 2a), we calculate its LPL induced BOPV conductivity. In practice, the BZ integration in Eq. (2) is $\int \frac{d^3k}{(2\pi)^3} = \frac{1}{V}\sum_k w_k$, where $V$ is the total volume of simulation supercell and $w_k$ is the weight of each $k$-point. In the 3D periodic boundary condition, the supercell of 2D materials contains artificial vacuum space along $z$, whose contribution needs to be eliminated. According to previous works, we rescale this result by using an effective thickness of 2D materials $d$ (taken to be 0.6 nm), which is estimated by the layer-to-layer distance when these 2D materials are van der Waals stacked into bulk. Thus, we can rescale the photo-conductivity by $\sigma^{2D} = \sigma^{SC} h/d$, where $\sigma^{SC}$ and $h$ are the supercell calculated conductivity and the supercell lattice constant along $z$, respectively.[43,44] This makes the second order conductivity of 2D materials consistent with conventional quantities of 3D bulk materials. In the following, we will report the $\sigma^{2D}$ values. As shown in Figure 2b, one sees that consistent with our previous symmetry analysis, $\sigma_{xx}^{y;L_z}$ and $\sigma_{yy}^{y;L_z}$ are exactly zero through all optical frequency, while $\sigma_{xx}^{x;L_z}$ and $\sigma_{yy}^{x;L_z}$ are finite. This numerical results demonstrate that the nonlinear BOPV current flows along the $x$-direction, which is normal to the mirror plane. Our BPV conductivity calculation confirms that the $x$-direction charge current is zero (see Supporting Information), which also agrees with previous works. Thus, it suggests that the same amount of electrons carrying opposite $z$-component angular momentum are transporting along $x$ and $-x$ directions, which can be called as pure orbital current (charge current $\mathcal{J}^x = e(v^{x;l_z} + v^{x;-l_z}) = 0$, orbital current $\mathcal{J}^{x;L_z} = l_z v^{x;l_z} - l_z v^{x;-l_z} \neq 0$, where $l_z$ is the averaged $z$-component angular momentum expectation value, similar as spin up and spin down in the spin DOF). As for the $y$-direction electron current, it is a pure charge current with $\mathcal{J}^y = e(v^{y;l_z} + v^{y;-l_z}) \neq 0$ and orbital current $\mathcal{J}^{y;L_z} = l_z v^{y;l_z} - l_z v^{y;-l_z} = 0$, indicating that all electrons are moving along the same direction ($y$ or $-y$) and the amount of electrons carrying $l_z$ and $-l_z$ are the same. This is consistent with illustrations in Figure 1a.

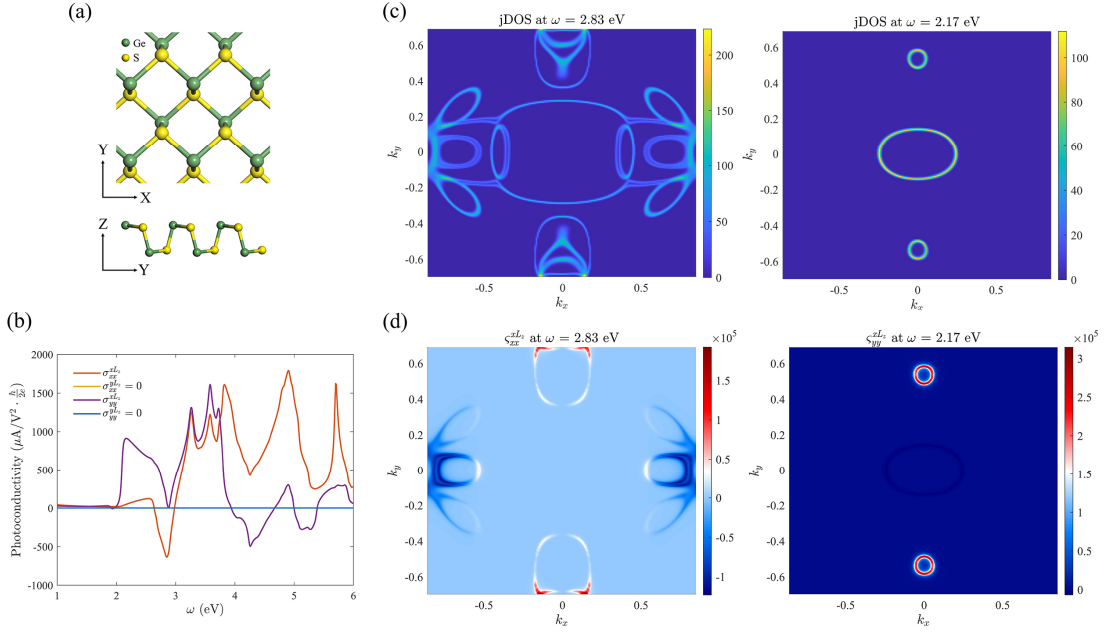

**Figure 2.** (a) Atomic geometry of monolayer GeS. (b) Calculated BOPV conductivity as a function of incident linearly polarized light frequency. (c) jDOS $\tilde{\rho}(\omega, \boldsymbol{k})$ distribution in the first BZ at $\omega = 2.83$ eV and $\omega = 2.17$ eV. (d) $\boldsymbol{k}$-resolved BOPV conductivity $\varsigma_{xx}^{x;L_z}$ (at $\omega = 2.83$ eV) and $\varsigma_{yy}^{x;L_z}$ (at $\omega = 2.17$ eV).

When the incident light energy is below the direct bandgap of monolayer GeS (1.91 eV), all photo-conductivities are zero since no interband transition occurs. At incident optical energy of $\hbar\omega = 2.83$ eV, the nonlinear BOPV conductivity $\sigma_{xx}^{x;L_z}$ reaches a negative peak of $-636.2 \frac{\mu A}{V^2} \frac{\hbar}{2e}$. Here the unit $\frac{\mu A}{V^2}$ is that of nonlinear charge BPV conductivity, and $\frac{\hbar}{2e}$ converts it to angular momentum, similar as that from Hall conductance to spin Hall conductance. If one measures magnetic moment, this unit becomes $g \frac{\mu A}{V^2} \frac{\mu_B}{2e}$, where $g$ is Landé g-factor ($g_L \simeq -1$, $g_S \simeq -2$) and $\mu_B$ is Bohr magneton. In order to further examine its momentum space contribution, we first plot the $\boldsymbol{k}$-resolved joint density of states (jDOS) $\tilde{\rho}(\omega, \boldsymbol{k})$ at $\omega = 2.83$ eV (Figure 2c, left panel). The jDOS reads

$$\rho_{cv}(\omega) = \frac{1}{(2\pi)^2} \int_{BZ} \tilde{\rho}(\omega, \boldsymbol{k}) d^2\boldsymbol{k} = \frac{1}{(2\pi)^2} \int_{BZ} \sum_{c,v} \delta(\varepsilon_{ck} - \varepsilon_{vk} - \hbar\omega) d^2\boldsymbol{k}, \quad (3)$$

where $\varepsilon_{nk}$ is eigenvalue of band-$n$ at momentum $\boldsymbol{k}$, and $c$ and $v$ represent the conduction and valence bands, respectively. The integral is taking in the first BZ. According to Sokhotski-Plemelj formula, jDOS represents the resonant band transition between band-$l$ and band-$n$ in Eq. (2). One could see that $\tilde{\rho}_{cv}(\omega = 2.83 \text{ eV}, \boldsymbol{k})$ is mainly contributed around the $\pm X$ and $\pm Y$ points in the BZ. We next plot the real part of the integrand of Eq. (2), $\varsigma_{xx}^{x;L_z}(\omega, \boldsymbol{k}) = \Re \sum_{lmn}^{\Omega=\pm\omega} f_{ln} \frac{v_{nl}^x v_{lm}^x j_{mn}^{x;L_z}}{(\omega_{nm}-i/\tau)(\omega_{nl}+\Omega-i/\tau)}$ (Figure 2d, left panel), integrating which over the first BZ yields $\chi_{xx}^{x;L_z} = \sigma_{xx}^{x;L_z}$. One clearly sees that in the momentum space, $\varsigma_{xx}^{x;L_z}(\omega, \boldsymbol{k})$ keeps the $\mathcal{M}_x$ symmetry, and is mainly contributed around $\pm X$ and $\pm Y$. The jDOS around $\Gamma$ point does not contribute any significant photo-conductivities. When the $y$-polarized LPL is shined, it reaches a peak of 913.95 $\frac{\mu A}{V^2}\frac{\hbar}{2e}$ at $\omega = 2.17$ eV. The $\boldsymbol{k}$-resolved jDOS and $\varsigma_{yy}^{x;L_z}(\omega, \boldsymbol{k})$ are shown in the right panels of Figures 2c and 2d. Again, the distribution shows $\mathcal{M}_x$ symmetry, which locates around the valleys at $(0, \pm 0.54, 0)$ Å$^{-1}$, but not around $\Gamma$.

The SOC interaction that breaks the spin rotational symmetry usually splits the spin up and spin down degeneracy in the centrosymmetric broken systems (such as Rashba and Dresselhaus splitting). Here we show that such spin polarization in band dispersion also produces finite BSPV effect. In Figure 3a we plot the calculated BSPV conductivity of monolayer GeS. Analogues to the BOPV, under LPL illumination, BSPV also flows along the $x$-direction, giving finite $\sigma_{xx}^{x;S_z}$ and $\sigma_{yy}^{x;S_z}$, while $\sigma_{xx}^{y;S_z} = \sigma_{yy}^{y;S_z} = 0$. However, we find that the magnitude of BSPV conductivities are generally much smaller than that of the BOPV. Comparing Figure 2b and 3a, one could see that the magnitude of BOPV conductivity is about one order of magnitude larger than the BSPV conductivity. For example, the $\sigma_{xx}^{x;S_z}(\omega = 2.83 \text{ eV}) = 113.37 \frac{\mu A}{V^2}\frac{\hbar}{2e}$ and $\sigma_{yy}^{x;S_z}(\omega = 2.17 \text{ eV}) = 3.60 \frac{\mu A}{V^2}\frac{\hbar}{2e}$, much smaller than the corresponding BOPV magnitudes at the same frequency. We also plot their momentum space contributions ($\varsigma_{xx}^{x;S_z}$ and $\varsigma_{yy}^{x;S_z}$, Figure 3b), which show that they locate similarly as in the BOPV conductivities, and the $\widehat{\mathcal{M}}_x$ symmetry still retains. Note that these 2D ferroelectric monolayers possess four electron valleys in the first BZ (near $\pm X$ and $\pm Y$). Hence, we show that the photocurrent is mainly contributed from these valleys, which may provide promising physical properties among orbitronics, spintronics, and valleytronics.

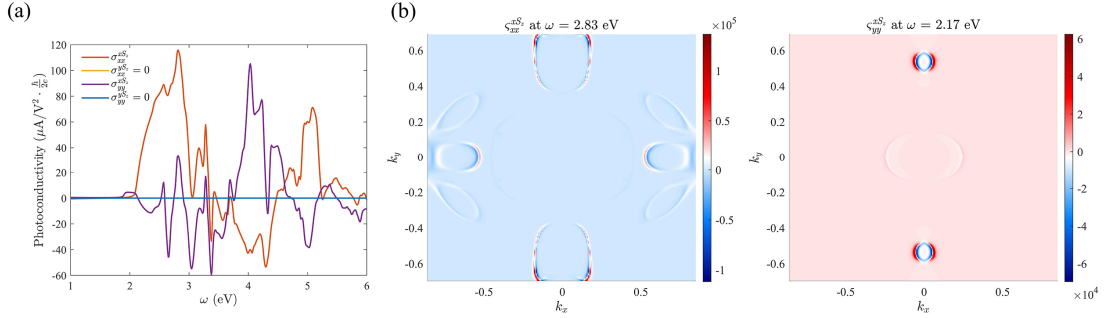

**Figure 3.** (a) BSPV photoconductivity component of monolayer GeS under LPL. (b) Momentum space distributions of the integrand $\varsigma_{xx}^{x;S_z}(\omega = 2.83 \text{ eV}, \boldsymbol{k})$ and $\varsigma_{yy}^{x;S_z}(\omega = 2.17 \text{ eV}, \boldsymbol{k})$.

According to solid state physics theory, orbital moment in a bulk material is usually strongly quenched by the symmetric crystal field, so that it is the spin polarization that mainly contributes to the total magnetic moment. Hence, the orbital moment contribution is omitted in most cases. However, here we find that BOPV conductivity is generally much larger than that of BSPV. According to Eq. (2), the dominate interband contribution is a two-band transition, namely, $|m\rangle = |n\rangle$, and the $|l\rangle$ band lies on the other side of the Fermi level (hence $f_{ln} \neq 0$). We will limit our discussion on this two-band model. Thus, the difference between BOPV and BSPV conductivity can be understood by comparing $\langle\{v_x, L_z\}\rangle$ and $\langle\{v_x, S_z\}\rangle$ for the low energy bands (near Fermi level), which is determined by the velocity and orbital/spin texture. In order to illustrate it more clearly, we plot the $\boldsymbol{k}$-space distribution of orbital and spin angular momentum ($\langle L_z \rangle$ and $\langle S_z \rangle$) of the highest valence band (VB) and the second highest valence band (VB−1), as shown in Figure 4a and Figure 4b. Here VB and VB−1 are actually Rashba splitting bands. One clearly observes that the $\langle L_z \rangle$ distribution on the VB and VB−1 are similar, while the $\langle S_z \rangle$ distribution on them show opposite values.

This is because that the orbital texture is determined by the crystal field once the material forms, and changes marginally under SOC. On the other hand, the Rashba-type spin splitting yields that $\langle S_z \rangle$ flips its sign between the VB and VB−1 at each $\boldsymbol{k}$. We also plot the spin and orbital angular momentum distributions of the lowest two conduction bands in Supporting Information, and similar results can be seen. The velocity texture distributions on VB and VB−1 are also similar (but not identical) (Figure 4c). We plot $\langle \{v_x, L_z\} \rangle$ and $\langle \{v_x, S_z\} \rangle$ of bands near the Fermi level in Supporting Information. From all these evidences, we show that the crystal field determined orbital responses are similar at the Rashba splitting bands, while their contributions to the spin responses are opposite (but not completely cancelled due to small velocity distribution difference). Therefore, the BSPV conductivity is usually much smaller than that of BOPV.

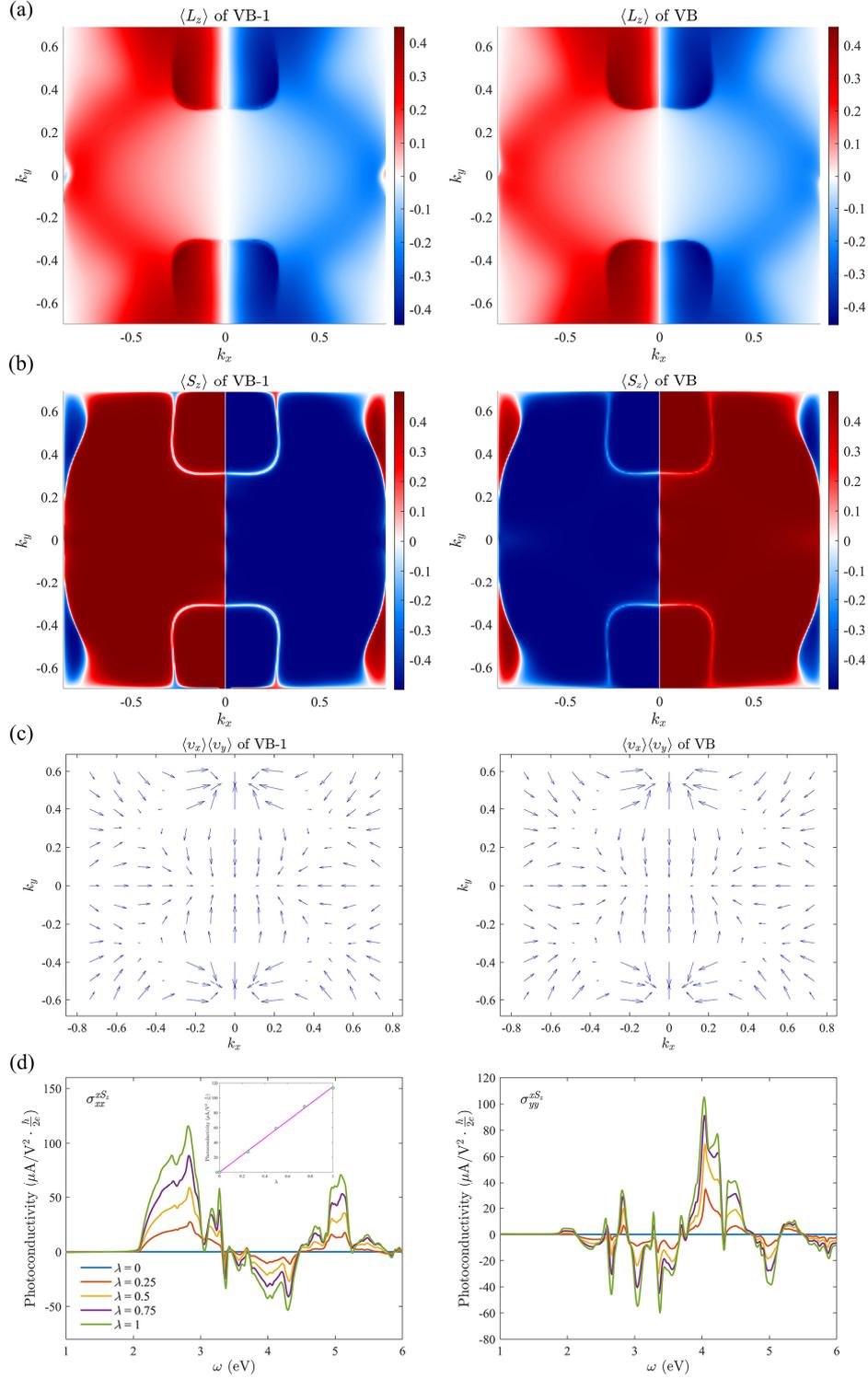

**Figure 4.** Momentum space distribution of (a) $\langle L_z \rangle$, (b) $\langle S_z \rangle$ and (c) velocity texture $(\langle v_x \rangle, \langle v_y \rangle)$ of VB and VB–1 in the first BZ. (d) BSPV response function under different SOC strength parameter $\lambda$. Inset: The peak magnitude change of BSPV $\sigma_{xx}^{x;S_z}(\omega = 2.83 \text{ eV})$ as a function of $\lambda$. Linear relation can be clearly seen. One has to note that the band dispersion will be significantly changed under very strong SOC,

so that such linearity may not hold when $\lambda$ is very big.

In order to further understand the mechanism of BOPV and BSPV photo-conductivities, we artificially tune the SOC interaction $H_{SO} = \alpha_{SO} \mathbf{S} \cdot \mathbf{L}/\hbar^2$ strength by multiplying a pre-factor $\lambda \in [0,1]$. Here $\lambda = 0$ turns off the SOC, and $\lambda = 1$ indicates full SOC. We find that the BOPV (and BPV) conductivity marginally changes under different $\lambda$ (see Supporting Information). However, the BSPV conductivity linearly reduces to zero from $\lambda = 1$ (full SOC) to $\lambda = 0$ (no SOC), as shown in Figure 4d. This clearly demonstrates that the BOPV effect is ubiquitous even without SOC since orbital texture originates from crystal field, while SOC is crucial for BSPV effect in these nonmagnetic systems. Therefore, we could conclude that the BOPV effect arises when crystal is formed, and then it leads to BSPV effect through a finite SOC interaction ($H_{SO} \propto \mathbf{S} \cdot \mathbf{L}$). Similar relation can also be seen in the orbital and spin Hall effects.[16] Note that very strong SOC may not necessarily imply further enhanced BSPV, as the band dispersion would be significantly affected.

For the ferroelectric materials, one could easily apply external (electrical, mechanical, and optical) fields to modulate its polarization (for example, from $P_0$ to $-P_0$). The transition barrier between different ferroic orders is usually a high symmetric geometry, which is centrosymmetric and not electrically polarized ($P = 0$). We now examine the BOPV and BSPV photo-conductivity under different electric polarizations. In Figure 5 we plot the polarization dependent BOPV and BSPV photo-conductivities. One clearly observes that all these conductivities diminishes at $P = 0$ state. This is consistent with symmetry analysis, $\hat{I} N_{abc;z}(\mathbf{k}) = -N_{abc;z}(-\mathbf{k})$, where $\hat{I}$ is inversion symmetry operator (angular momenta $L$ and $S$ are invariant under $\hat{I}$). We also note that when the polarization flips (corresponding to a 180°-rotation from $P_0$ to $-P_0$), the photo-conductivities reverse their flowing direction while keeping same magnitudes. If a 90°-rotation occurs, these conductivities also rotate 90°, flowing along the $\pm y$-direction. Hence, one could control the ferroic polarization order to manipulate the BOPV and BSPV photocurrents, as well as the BPV effect.

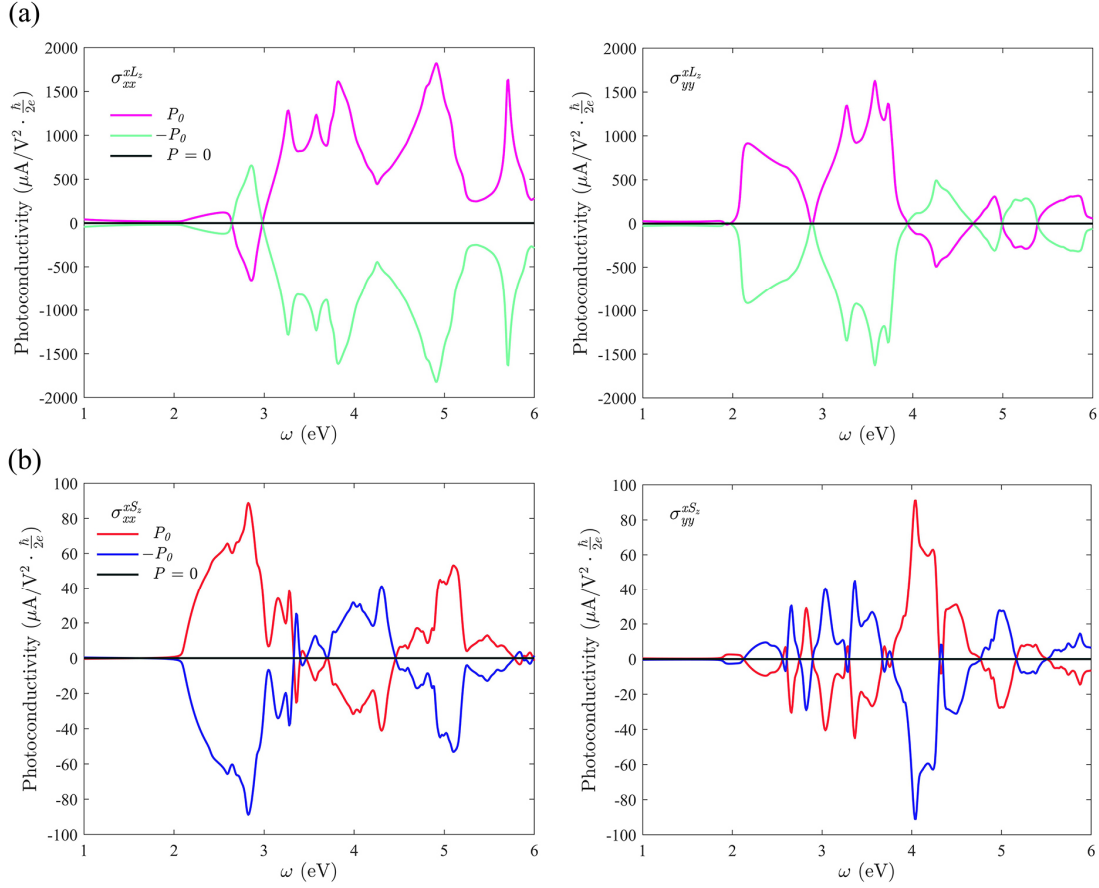

**Figure 5.** Polarization dependent (a) BOPV and (b) BSPV conductivity of monolayer GeS. The reversal of polarization $P$ flips the photocurrent, while the high symmetric structure ($P = 0$) forbids any photocurrents.

We now calculate the BOPV and BSPV conductivities for other analogues, namely, monolayers GeSe, SnS, SnSe, GeTe, SnTe, and Bi. Note that even though the monolayer Bi is a single elemental material, Peierls instability occurs due to strong s and p orbital hybridization, which leads to charge transfer within each atomic layer. Thus, the monolayer Bi also shows in-plane ferroelectricity and fascinating optical properties. All these BOPV and BSPV photo-conductivity results are shown in Figure 6. For the BOPV conductivity, we observe clear similarities for all these systems, because their electronic band structure can be described by the same low energy model.[30] By comparing the main peaks in Figure 6a and 2b, we find that (for the group IV-VI systems) when the system is composed by small cation and large anion, the BOPV photo-conductivity shows larger peak values (over $10,000 \ \frac{\mu A}{V^2}\frac{\hbar}{2e}$). Hence, the monolayer GeTe shows largest photocurrent responses, while the orbital photoconductivity of monolayer SnS is smallest. However, the BSPV does not have such similarity as the SOC interaction strength (proportional to $Z^4$) determines its responses.

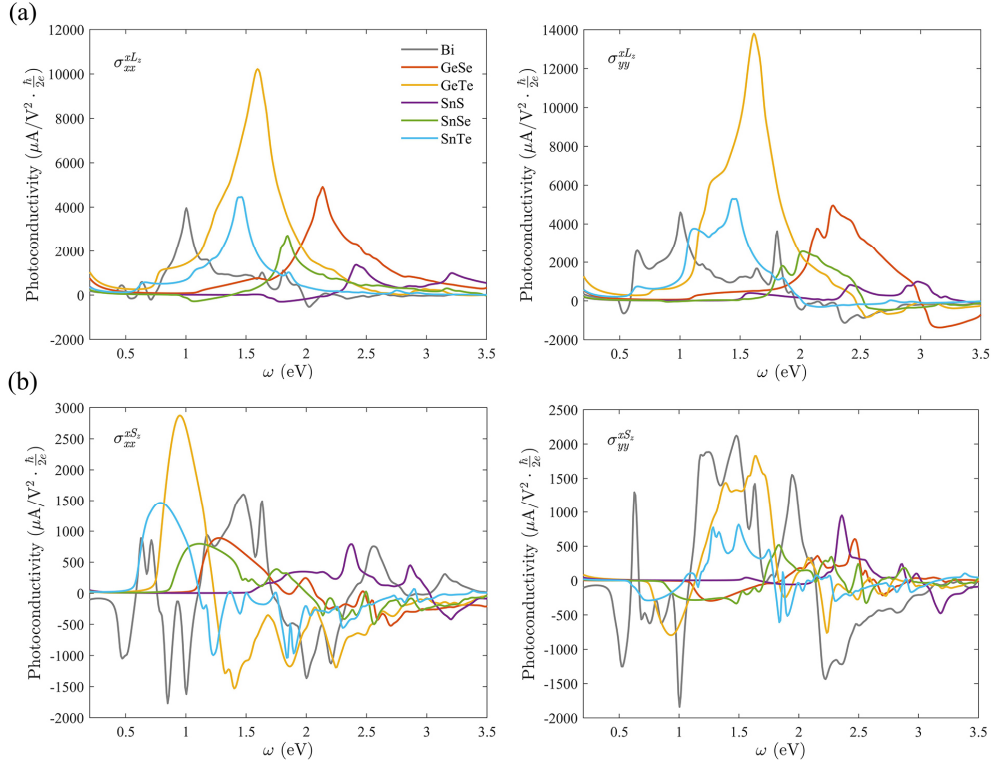

**Figure 6.** (a) BOPV and (b) BSPV conductivity coefficients of other similar 2D ferroelectric monolayers.

**Table 1.** Polarization dependent bulk (charge, spin, and orbital) photovoltaic conductivities along $x$. Here $\eta$ represents left- or right-handed CPL. Polarization of LPL is along $x$ or $y$. All eight different types of photocurrents can be realized under light and ferroicity. Symbols SC and IC indicate shift and injection current, respectively. For the $y$-directional currents, one can apply a 90°-rotation to yield similar results.

| Polarization | $P_{+y}$ | $P_{-y}$ | $P_{+x}$ | $P_{-x}$ |
|---|---|---|---|---|
| Mirror | $\widehat{\mathcal{M}}_x$ | $\widehat{\mathcal{M}}_x$ | $\widehat{\mathcal{M}}_y$ | $\widehat{\mathcal{M}}_y$ |
| LPL | $\sigma_{\text{LPL}}^{x;O_z}$ (IC) | $-\sigma_{\text{LPL}}^{x;O_z}$ (IC) | $\sigma_{\text{LPL}}^{x}$ (SC) | $-\sigma_{\text{LPL}}^{x}$ (SC) |
| CPL | $\sigma_\eta^x = -\sigma_{-\eta}^x$ (IC) | $-\sigma_\eta^x = \sigma_{-\eta}^x$ (IC) | $\sigma_\eta^{x;O_z} = -\sigma_{-\eta}^{x;O_z}$ (SC) | $-\sigma_\eta^{x;O_z} = \sigma_{-\eta}^{x;O_z}$ (SC) |

**Conclusion.** We predict robust and pure bulk photovoltaic currents in the carrier orbital and spin degrees of freedom. Using nonmagnetic 2D ferroelectric materials (GeS and its analogues) as exemplary materials, we show that the mirror symmetry forbidden BPV conductivity actually contains hidden electron motions, which carries orbital moment flow with zero net electric charge current. Under SOC interaction, the photo-induced orbital current could convert into spin current. Both of these currents are perpendicular to conventional BPV electric current, so that they can be purely and exclusively detected and used. When ferroic order switches, the photo-conductivities rotate their directions accordingly. We summarize such polarization and light dependent photovoltaic effects in Table 1. Our prediction of pure BOPV and BSPV effects can be easily detected and observed in experiments, and may provide potential ultrafast spintronic and orbitronic applications of 2D in-plane ferroelectric materials, in addition to their electronic features, especially when a four-terminal device is applied.

**Methods.** We use first-principles density functional theory to calculate the geometric, electronic, and optical properties of 2D monolayer GeS and analogous systems, as implemented in the Vienna *ab initio* simulation package (VASP).[45] The generalized gradient approximation (GGA) in the Perdew-Burke-Ernzerhof (PBE) form[46] is adopted to treat the exchange correlation functional in the Kohn-Sham equation. A vacuum space of 12 Å along the out-of-plane $z$-direction is used, to eliminate the interactions between different periodic images. Projector-augmented wave (PAW)[47] method is used to treat the core electrons, and the valence electrons are represented by planewave basis set, with a kinetic cutoff energy chosen to be 350 eV. The first Brillouin zone is represented by the Monkhorst-Pack $k$-mesh scheme[48] with a (9×9×1) grid for geometric and electronic structure calculations. Convergence criteria of total energy and force component on each ion are set as $1×10^{-7}$ eV and 0.01 eV/Å, respectively. Spin-orbit coupling interactions are self-consistently included in all calculations, unless otherwise noted. In order to evaluate the nonlinear optical conductivities, we fit the electronic structure by atomic orbital tight-binding model in atomic orbital basis set (s and p orbitals), as implemented in the Wannier90 package,[49,50] and the optical conductivities are integrated on a denser $k$-mesh of (901×901×1) grid. The convergence of $k$-grid density is carefully tested. As for the estimate of orbital angular momentum contributed intra-atomically, we use $|s\rangle = Y_0^0$, $|p_x\rangle = \frac{1}{\sqrt{2}}(Y_1^{-1} - Y_1^1)$, $|p_y\rangle = \frac{i}{\sqrt{2}}(Y_1^{-1} + Y_1^1)$, and $|p_z\rangle = Y_1^0$ as basis set and calculate their matrix components $\langle m|L_z|n\rangle$, with the intra-atomic orbital angular momentum $L_z = i\hbar \begin{pmatrix} 0 & -\frac{1}{2}\sigma_- \\ \frac{1}{2}\sigma_+ & 0 \end{pmatrix}$, where $\sigma_\pm = \sigma_x \pm i\sigma_y$. The spin operators are proportional to conventional Pauli matrix, $S_i = \frac{\hbar}{2}\sigma_i$, where $\sigma_i$ ($i = x, y, z$) are Pauli matrices. We test and verify our calculation procedure by comparing with previous orbital momentum calculations and BPV effect computations.

**Acknowledgments.** This work was supported by the National Natural Science Foundation of China under Grant Nos. 21903063 and 11974270, and the Startup Funding Program of Xi'an Jiaotong University. J.Z. acknowledges helpful discussions with Haowei Xu and Dr. Hua Wang, Dr. Ruixiang Fei, and Prof. Yang Gao on the nonlinear optical effect and theory, and discussions with Prof. Yi Pan for potential observations.

**Supplementary Information.** Electronic band dispersion of monolayer GeS and its tight binding fitting, BPV and BOPV photo-conductivities of monolayer GeS under different SOC strengths, $k$-space distribution of spin $\langle S_z\rangle$ and orbital angular momentum $\langle L_z\rangle$ of the CB and CB+1, momentum space resolved $\langle\{v_x, S_z\}\rangle$ and $\langle\{v_x, L_z\}\rangle$ for the bands near Fermi level, circularly polarized light induced BPV, BOPV, and BSPV conductivities, and BPV photo-conductivity coefficient of monolayer GeS.

**Corresponding authors**: *J.Z.: jianzhou@xjtu.edu.cn